\documentstyle[12pt,epsf]{article}
\begin{document}
\newcommand {\ee}{\end{equation}}
\newcommand {\bea}{\begin{eqnarray}}
\newcommand {\eea}{\end{eqnarray}}
\newcommand {\nn}{\nonumber \\}
\newcommand {\Tr}{{\rm Tr\,}}
\newcommand {\tr}{{\rm tr\,}}
\newcommand {\e}{{\rm e}}
\newcommand {\etal}{{\it et al.}}
\newcommand {\m}{\mu}
\newcommand {\n}{\nu}
\newcommand {\pl}{\partial}
\newcommand {\p} {\phi}
\newcommand {\vp}{\varphi}
\newcommand {\vpc}{\varphi_c}
\newcommand {\al}{\alpha}
\newcommand {\be}{\beta}
\newcommand {\ga}{\gamma}
\newcommand {\Ga}{\Gamma}
\newcommand {\x}{\xi}
\newcommand {\ka}{\kappa}
\newcommand {\la}{\lambda}
\newcommand {\La}{\Lambda}
\newcommand {\si}{\sigma}
\newcommand {\th}{\theta}
\newcommand {\Th}{\Theta}
\newcommand {\om}{\omega}
\newcommand {\Om}{\Omega}
\newcommand {\ep}{\epsilon}
\newcommand {\vep}{\varepsilon}
\newcommand {\na}{\nabla}
\newcommand {\del}  {\delta}
\newcommand {\Del}  {\Delta}
\newcommand {\mn}{{\mu\nu}}
\newcommand {\ls}   {{\lambda\sigma}}
\newcommand {\ab}   {{\alpha\beta}}
\newcommand {\half}{ {\frac{1}{2}} }
\newcommand {\fourth} {\frac{1}{4} }
\newcommand {\sixth} {\frac{1}{6} }
\newcommand {\sqg} {\sqrt{g}}
\newcommand {\fg}  {\sqrt[4]{g}}
\newcommand {\invfg}  {\frac{1}{\sqrt[4]{g}}}
\newcommand {\sqZ} {\sqrt{Z}}
\newcommand {\gbar}{\bar{g}}
\newcommand {\sqk} {\sqrt{\kappa}}
\newcommand {\sqt} {\sqrt{t}}
\newcommand {\reg} {\frac{1}{\epsilon}}
\newcommand {\fpisq} {(4\pi)^2}
\newcommand {\Lcal}{{\cal L}}
\newcommand {\Ocal}{{\cal O}}
\newcommand {\Dcal}{{\cal D}}
\newcommand {\Ncal}{{\cal N}}
\newcommand {\Mcal}{{\cal M}}
\newcommand {\scal}{{\cal s}}
\newcommand {\Dvec}{{\hat D}}   
\newcommand {\dvec}{{\vec d}}
\newcommand {\Evec}{{\vec E}}
\newcommand {\Hvec}{{\vec H}}
\newcommand {\Vvec}{{\vec V}}
\newcommand {\Btil}{{\tilde B}}
\newcommand {\ctil}{{\tilde c}}
\newcommand {\Stil}{{\tilde S}}
\newcommand {\Ztil}{{\tilde Z}}
\newcommand {\altil}{{\tilde \alpha}}
\newcommand {\betil}{{\tilde \beta}}
\newcommand {\latil}{{\tilde \lambda}}
\newcommand {\ptil}{{\tilde \phi}}
\newcommand {\Ptil}{{\tilde \Phi}}
\newcommand {\natil} {{\tilde \nabla}}
\newcommand {\ttil} {{\tilde t}}
\newcommand {\Dhat}{{\hat D}}
\newcommand {\Shat}{{\hat S}}
\newcommand {\shat}{{\hat s}}
\newcommand {\xhat}{{\hat x}}
\newcommand {\Zhat}{{\hat Z}}
\newcommand {\Gahat}{{\hat \Gamma}}
\newcommand {\nah} {{\hat \nabla}}
\newcommand {\gh}  {{\hat g}}
\newcommand {\labar}{{\bar \lambda}}
\newcommand {\cbar}{{\bar c}}
\newcommand {\bbar}{{\bar b}}
\newcommand {\Bbar}{{\bar B}}
\newcommand {\psibar}{{\bar \psi}}
\newcommand {\chibar}{{\bar \chi}}
\newcommand {\bbartil}{{\tilde {\bar b}}}
\newcommand {\intfx} {{\int d^4x}}
\newcommand {\inttx} {{\int d^2x}}
\newcommand {\change} {\leftrightarrow}
\newcommand {\ra} {\rightarrow}
\newcommand {\larrow} {\leftarrow}
\newcommand {\ul}   {\underline}
\newcommand {\pr}   {{\quad .}}
\newcommand {\com}  {{\quad ,}}
\newcommand {\q}    {\quad}
\newcommand {\qq}   {\quad\quad}
\newcommand {\qqq}   {\quad\quad\quad}
\newcommand {\qqqq}   {\quad\quad\quad\quad}
\newcommand {\qqqqq}   {\quad\quad\quad\quad\quad}
\newcommand {\qqqqqq}   {\quad\quad\quad\quad\quad\quad}
\newcommand {\qqqqqqq}   {\quad\quad\quad\quad\quad\quad\quad}
\newcommand {\lb}    {\linebreak}
\newcommand {\nl}    {\newline}

\newcommand {\vs}[1]  { \vspace*{#1 cm} }

\newcommand {\MPL}  {Mod.Phys.Lett.}
\newcommand {\NP}   {Nucl.Phys.}
\newcommand {\PL}   {Phys.Lett.}
\newcommand {\PR}   {Phys.Rev.}
\newcommand {\PRL}   {Phys.Rev.Lett.}
\newcommand {\CMP}  {Commun.Math.Phys.}
\newcommand {\AP}   {Ann.of Phys.}
\newcommand {\PTP}  {Prog.Theor.Phys.}
\newcommand {\NC}   {Nuovo Cim.}
\newcommand {\CQG}  {Class.Quantum.Grav.}


\font\smallr=cmr5
\def\ocirc#1{#1^{^{{\hbox{\smallr\llap{o}}}}}}
\def\ogamma{\ocirc{\gamma}{}}
\def\oM{{\buildrel {\hbox{\smallr{o}}} \over M}}
\def\osigma{\ocirc{\sigma}{}}

\def\overleftrightarrow#1{\vbox{\ialign{##\crcr
 $\leftrightarrow$\crcr\noalign{\kern-1pt\nointerlineskip}
 $\hfil\displaystyle{#1}\hfil$\crcr}}}
\def\overnab{{\overleftrightarrow\nabslash}}

\def\va{{a}}
\def\vb{{b}}
\def\vc{{c}}
\def\tilpsi{{\tilde\psi}}
\def\tbpsi{{\tilde{\bar\psi}}}

\def\Dslash{{}\hbox{\hskip2pt\vtop
 {\baselineskip23pt\hbox{}\vskip-24pt\hbox{/}}
 \hskip-11.5pt $D$}}
\def\nabslash{{}\hbox{\hskip2pt\vtop
 {\baselineskip23pt\hbox{}\vskip-24pt\hbox{/}}
 \hskip-11.5pt $\nabla$}}
\def\xislash{{}\hbox{\hskip2pt\vtop
 {\baselineskip23pt\hbox{}\vskip-24pt\hbox{/}}
 \hskip-11.5pt $\xi$}}
\def\leftnabla{{\overleftarrow\nabla}}

\def\delL{{\delta_{LL}}}
\def\delG{{\delta_{G}}}
\def\delc{{\delta_{cov}}}

\newcommand {\gago}  {\gamma_5}
\newcommand {\Ktil}  {{\tilde K}}
\newcommand {\Ltil}  {{\tilde L}}
\newcommand {\Kbar}  {{\bar K}}
\newcommand {\GfMp}  {G^{5M}_+}
\newcommand {\GfMpm}  {G^{5M'}_-}
\newcommand {\GfMm}  {G^{5M}_-}
\def\Aslash{{}\hbox{\hskip2pt\vtop
 {\baselineskip23pt\hbox{}\vskip-24pt\hbox{/}}
 \hskip-11.5pt $A$}}
\def\kslash{{}\hbox{\hskip2pt\vtop
 {\baselineskip23pt\hbox{}\vskip-24pt\hbox{/}}
 \hskip-8.5pt $k$}}@@
\def\dbslash{{}\hbox{\hskip2pt\vtop
 {\baselineskip23pt\hbox{}\vskip-24pt\hbox{$\backslash$}}
 \hskip-11.5pt $\partial$}}
\def\Kbslash{{}\hbox{\hskip2pt\vtop
 {\baselineskip23pt\hbox{}\vskip-24pt\hbox{$\backslash$}}
 \hskip-11.5pt $K$}}
\def\Ktilbslash{{}\hbox{\hskip2pt\vtop
 {\baselineskip23pt\hbox{}\vskip-24pt\hbox{$\backslash$}}
 \hskip-11.5pt ${\tilde K}$}}
\def\Ltilbslash{{}\hbox{\hskip2pt\vtop
 {\baselineskip23pt\hbox{}\vskip-24pt\hbox{$\backslash$}}
 \hskip-11.5pt ${\tilde L}$}}
\def\Kbarbslash{{}\hbox{\hskip2pt\vtop
 {\baselineskip23pt\hbox{}\vskip-24pt\hbox{$\backslash$}}
 \hskip-11.5pt ${\bar K}$}}
\def\Lbarbslash{{}\hbox{\hskip2pt\vtop
 {\baselineskip23pt\hbox{}\vskip-24pt\hbox{$\backslash$}}
 \hskip-11.5pt ${\bar L}$}}

\begin{flushright}
Nov. 1998; June 1999 rev.\\
hep-th/9811094 \\
BNL-preprint\\
US-preprint,US-98-09
\end{flushright}

\vspace{0.5cm}

\begin{center}
{\Large\bf 
Temperature in Fermion Systems\\
 and \\
the Chiral Fermion Determinant }

\vspace{1.5cm}
{\large Shoichi ICHINOSE
          \footnote{
On leave of absence from 
Department of Physics, University of Shizuoka,
Yada 52-1, Shizuoka 422-8526, Japan (Address after Nov.6, 1998).\nl          
E-mail address:\ ichinose@u-shizuoka-ken.ac.jp
                  }
}
\vspace{1cm}

{\large Physics Department, Brookhaven National Laboratory,\\
Upton, NY 11973, USA}


\end{center}
\vfill
{\large Abstract}\nl
We give an interpretation to the issue of
the chiral determinant in the heat-kernel
approach. The extra dimension (5-th dimension)
is interpreted as (inverse) 
temperature. The 1+4 dim
Dirac equation is naturally derived by the Wick
rotation for the temperature. In order to define
a ``good'' temperature, we 
choose those solutions of the Dirac equation which
propagate in a {\it fixed direction} in the extra coordinate.
This choice fixes the regularization of the fermion determinant. 
The 1+4 dimensional
Dirac mass ($M$) is naturally introduced and the relation:\ 
$|$4 dim electron momentum$|$\ $\ll$ $|M|$ $\ll$ ultraviolet cut-off, 
naturally appears.
The chiral anomaly is explicitly derived for the 2 dim
Abelian model. Typically two different regularizations 
appear depending on the choice of propagators.
One corresponds to the chiral theory, the other
to the non-chiral (hermitian) theory.

\vspace{0.5cm}
PACS NO:\ 11.30.Rd,11.25.Db,05.70.-a,11.10.Kk,11.10.Wx \nl

\newpage
\section{Introduction}
Since Dirac introduced the first-order differential operator
(Dirac operator) to describe the fermion propagation,
we have been provided by a tool to describe rich fermion physics.
The wonderful success of QED shows its deep importance.
It is, however, true that interesting ambiguities related to the
subtlety in its vacuum structure remain with
the chiral anomaly problem.
In order to define the quantum field theory, we must generally
regularize the space-time. Equivalently, we must define
the measure for quantum fluctuations.
One of the fundamental regularizations is the lattice.
Since its birth, the lattice approach has suffered from 
the fermion problem known as species doubling.
In the necessity of simulating fermions on a lattice, 
a new regularization was found by \cite{Kap92,Jan92}
using an idea of the higher dimensional theory\cite{CH85}.
This approach is called ``Domain wall fermions,'' and 
has been well developed by some people
\cite{Sha93,CH94,Creu95,FS95,BS97PR,BS97PRL}.
Especially Ref.\cite{Sha93} made seminal development in the early stage.
The most characteristic point of the regularization is the
introduction of one extra dimension. In accord with this,
a new regularization mass parameter, that is, 
the 5 dimensional (dim) fermion
mass $M$, appears.
It controls the selection of chirality and plays the central
role in this regularization scheme.

A continuum analysis 
has also been done by Narayanan-Neuberger
and Randjbar-Daemi-Strathdee\cite{NN94,NN95,DS95NP,DS95PL,DS97PL}
in the form of the ``Overlap formalism''. 
Here we present another continuum formalism based on the heat-kernel.
In this approach, a ``temperature'' parameter $t$ is used.
Previously for the elliptic operator, $t$ is naturally introduced
because of the positivity of its eigenvalues.
For the Dirac operator, however, the direct application is difficult
because of the negative eigenvalues. 
( So far, in order to avoid the
difficulty, a roundabout method has customarily been taken, that is, we
transform the problem of $\det \Dslash$ into that of $\det {\Dslash}^2$. 
See, for example, Chapter IV of \cite{AAR91}.)
We solve this difficulty
by an idea based on the fact: when the temperature is defined
in a system, the system evolves in a {\it fixed direction}.

The present formalism is a continuum approach based on
the heat-kernel method. This is, of course, different
from the lattice discrete regularization approach.
Both approaches, however, have a common property:\ 
they are coordinate-based regularizations and are
characterized by length parameters, $t$ (inverse temperature or proper time)
and $a$ (lattice constant) respectively. Because of this we can
expect to find some common qualitative behaviours.
In fact the analysis presented in the following has been
developed by looking at the lattice formulation.
Conversely some qualitative features 
( such as the "wall" configulation and
the restriction to the regularization parameter $M$),
which are rather complicated to show in lattice, can be simply shown in the
present formalism. 

In the (continuum) field theory, we must generally treat the space-time
in some regularized way in order to control divergent quantities.
So far many regularization methods have been invented. They are,
with their characteristic parameters,  
Pauli-Vilars ($M$:\ heavy field
mass), dimensional ( $1/(n-4)$, $n$:\ space-time dimension ),
heat-kernel ($t$:\ inverse temperature), etc.. In this general
standpoint of regularization, what status the domain wall regularization
has ? Undoubtedly it has an advantageous point in regularizing
chiral quantities. But how does it essentially work ? 
It has the 1+4 dim fermion
mass $M$ as the characteristic regularization parameter. In the lattice
numerical simulation, 
we know the choice of $M$ is rather delicate. It should
be most appropriately chosen for a best-fit output. Why is 
the domain wall parameter $M$ 
so different from that in other familiar regularizations ? 
In order to understand these things qualitatively, we are urged to
have a new formalism which can be compared with known regularizations.

The main points found in this formalism are as follows.
\begin{enumerate}
\item
Different regularizations appear depending on
the choice of solutions in the 1+4 dim Dirac equation.
Using these regularizations, 
the Adler-Bell-Jackiw (U(1) chiral) anomaly is obtained
for 2 dim QED. 
Typically two regularizations appear. One gives
the anomaly of QED, the other gives that of the chiral QED. 
The ABJ anomaly, in the original overlap formalism, 
was obtained in \cite{NN94} .
\item
As the extra axis, it should be a half line ( not a line ) like the
temperature. ( Otherwise the infra-red cut-off appears in the chiral
anomaly, which should be avoided.)  This is in close analogy to the
domain wall and overlap formalisms\cite{Sha93}.
\item
The (+)- and ($-$)-domains are introduced simply as the
result of the property of the continuum Dirac operator $\Dvec$:\ 
$\Dvec\gago+\gago\Dvec=0$. This is contrasted with
the original formulation of the overlap where the sign
of 5 dim fermion masses distinguishes the two domains.
\item
The reason why the ``overlap'' of $|+>$ and $|->$ should
be taken in the anomaly calculation is manifest. 
The ``overlap''
in the partition function corresponds to a ``difference''
in the effective action.
\item
The characteristic limit of the present regularization
:\ $|k^\m|\ll |M|\ll 1/t$, is explained carefully.
\end{enumerate}

In Sec.2 we review the heat-kernel approach taking the 2 dim
quantum gravity as an example. 
It is the ordinary treatment
and should be compared with the proposed new one in the
remaining sections. We present the new treatment for
fermion systems in Sec.3. It is regarded as new regularization
(for the chiral determinant ) and is composed of three stages.
We apply it to the chiral anomaly calculation of 2 dim QED in Sec.4.
We conclude in Sec.5.  
\section{Heat-Kernel Approach}

The heat-kernel method has a long history as a tool
to define $\det \Dvec$ for a differential operator
$\Dvec$. 
Here the cap symbol ($\hat{\mbox{ }}$) denotes the operator 
nature of $\Dvec$.
Let us briefly review the properties necessary
for the next section. We take, as an explicit model, 
the 2 dim (Euclidean) quantum
scalar theory $\p$ on the background curved space $g_\mn$:\ 
\begin{eqnarray}
\Lcal[g_\mn,\p]=\sqg \p (-\half\na^2)\p\equiv\half\ptil {\Dvec}\ptil(x)
\ ,\ 
{\Dvec}\equiv -\fg\na_x^2 \invfg\ ,\ 
\ptil\equiv \fg~\p\ .
\label{HK1}
\end{eqnarray}
We focus on the partition function and the Weyl anomaly of this theory.
$\Dvec$ is a positive (semi)definite operator. The partition
function is expressed as 
\begin{eqnarray}
Z[g]=\int\Dcal\ptil\,\e^{-\inttx \Lcal}=(\det\Dvec)^{-\half}\nn
=\exp\{-\half\Tr\ln\Dvec\}=\exp[\half\Tr
\int_0^\infty\frac{\e^{-t\Dvec}}{t}dt+\mbox{const}\,]
\label{HK2}
\end{eqnarray}
Here we have used a useful formula for a matrix $A$:\ $
\int_0^\infty\frac{\e^{-t}-\e^{-tA}}{t}dt=\ln A
$.
This is well-defined when all eigenvalues of $A$ are positive\cite{foot1}. 
The heat-kernel is defined by
\begin{eqnarray}
G(x,y;t)\equiv <x|e^{-t\Dvec}|y>\com                   \nn
(\frac{\pl}{\pl t}+\Dvec_x)G(x,y;t)=0 \com \q
\lim_{t\ra +0}G(x,y;t)=\del^2(x-y)\com            \label{HK4}
\end{eqnarray}
where $<x|$ and $|y>$ 
are in the x-representation of $\Dvec$ (Dirac's bra-
and ket-vectors respectively). 
( This formalism is familiar in statistical mechanics
as the ``density matrix''\cite{Fey72}. )
The final equation is the boundary
condition we take.
The parameter $t$ has two properties:\ 
1)\ Dimension of $t$ is (length)$^2$ ($\Dvec$ is the quadratic
differential operator.);\ 
2)\ $t$ is restricted to be positive. 
The first property shows that the distribution in x-space at $t$ is
localized within a distance $\sqrt{t}$. The second one reflects
the property of heat:\ heat propagates from the high temperature
to the low temperature ( the second law of the thermodynamics). 
For the positive definite operator $\Dvec$, like the present
case, we can naturally
introduce the temperature as $1/t$ using this heat-kernel.

$G(x,y;t)$ is solved by the (weak-field) perturbation:
$g_\mn=\del_\mn+h_\mn, |h_\mn|\ll 1, \Dvec=-\pl^2-\Vvec,
\Vvec=-h_\mn\pl_\m\pl_\n-\pl_\n h_\mn \pl_\m-\fourth\pl^2h+O(h^2)$
(see Ref.\cite{II98}) using two ingredients.\nl
\flushleft{(i)\ Heat Equation without source(Free Equation)}\nl
\begin{eqnarray}
(\frac{\pl}{\pl t}-\pl^2)G_0(x-y;t)=0\com\q t>0\com
\lim_{t\ra +0}~G_0(x-y;t)=\del^2(x-y)\com\nn
G_0(x-y;t)=\int\frac{d^2k}{(2\pi)^2}exp\{-k^2t+ik\cdot (x-y)\}
=\frac{1}{4\pi t}e^{-\frac{(x-y)^2}{4t}}
\pr
\label{HK5}
\end{eqnarray}

\flushleft{(ii)\ Heat Propagator}\nl
\begin{eqnarray}
(\frac{\pl}{\pl t}-\pl^2)S(x-y;t-s)=\del(t-s)\del^2(x-y)\com
\lim_{t\ra+0}S(x-y;t)=\del^2(x-y)\com
\nn
S(x-y;t)=\int\frac{d^2k}{(2\pi)^2}\frac{dk^0}{2\pi}
\frac{exp\{-ik^0t+ik\cdot(x-y)\} }{-ik^0+k^2} 
=\th(t)G_0(x-y;t)
\pr
                                     \label{HK6}
\end{eqnarray}
We have imposed an appropriate boundary condition above,
corresponding to the one in (\ref{HK4}).
The perturbative solution is given by iterating
\begin{eqnarray}
G(x,y;t)=G_0(x-y;t)+\int d^2z\int^\infty_{-\infty}ds~
S(x-z;t-s){\hat V}(z)G(z,y;s)               \com
\label{f.15}
\end{eqnarray}
where $\Dvec=-\pl^2-{\hat V}$.
\q The lowest order solution of the present 2 dim scalar-gravity theory
 is given by
\begin{eqnarray}
G(x,x;t) &=& G_0(0;t)+G_1(x,x;t)+O(h^2)\nn
 &=& \frac{1}{4\pi t}(1+\frac{h}{2})-\frac{1}{24\pi}(\pl^2h-\pl_\m\pl_\n h_\mn)
                                  +O(h^2)+O(t)\label{HK8}\\
 &=& \frac{1}{4\pi t}(\sqg+O(h^2))-\frac{\sqg}{24\pi}(R+O(h^2))+O(t)\com
\end{eqnarray}
where the Riemann scalar curvature $R$ is expanded as
$R=\pl^2h_{\m\m}-\pl_\m\pl_\n h_\mn +O(h^2)$. 
Introducing an (inverse) ultra-violet cut-off $\ep$($\ra +0$) for $t$-integral
in (\ref{HK2}), we can
regularize $\Ga[g]\equiv \ln Z[g]$ as
\begin{eqnarray}
\Ga^{\mbox{reg}}[g]=\half \int_{\ep}^{\infty}\frac{dt}{t}
\inttx <x|\e^{-t\Dvec}|x>+\mbox{const}\pr
\label{HK9}
\end{eqnarray}
Its variation under the scale transformation $\del\ep=\Del\cdot\ep$
gives us the Weyl anomaly.
\begin{eqnarray}
\del\Ga^{\mbox{reg}}
=\half\left[ \inttx\,G(x,x;\ep)\right]\cdot\Del\nn
=\Del\cdot\inttx\left[ -\frac{1}{48\pi}\sqg R+\mbox{cosmological term}\right]
\pr
\label{HK10}
\end{eqnarray}
The same result is obtained by the measure change\cite{KF79,II96,II98}.
\begin{eqnarray}
\del\ln J=\del(\ln\,\det\frac{\pl\ptil'}{\pl\ptil})=-\Tr\al(x)\del^2(x-y)
=-\lim_{t\ra +0}\Tr\,\al(x)G(x,y;t)
\com
\label{HK10B}
\end{eqnarray}
where $\ptil'=\e^{-\al(x)}\ptil$.
This approach will be taken in the evaluation of the chiral anomaly
in Sec.4.
The result (\ref{HK10}) 
was exploited in the 2 dim quantum gravity\cite{Pol87}.

\q We pointed out (in eq.(\ref{HK6})) 
that the heat propagator $S(x-y;t)$ has the 
characteristic form\ :\ 
the theta function, $\th(t)$, times the free solution $G_0(x,y;t)$.
The theta function guarantees that the propagation is going
in the fixed direction, that is, forward in the $t$-axis.
We will utilize this fact in the choice of boundary conditions.

\section{Chiral Fermion Determinant}

Let us start with 4 dim Euclidean {\it massless} {\it Dirac} fermion $\psi$
in the external Abelian gauge field $A_\m$.\cite{foot2}
\begin{eqnarray}
\Lcal=\psibar\Dvec\psi\com\q
\Dvec=i\ga_\m(\pl_\m+i\e A_\m)\com\q
\Dvec^\dag=\Dvec\pr
\label{CF1}
\end{eqnarray}
We can formally express the determinant as in the previous section.
\begin{eqnarray}
Z[A]=\int\Dcal\psibar\Dcal\psi\e^{-\intfx\Lcal}=(\det\Dvec)^{+1}\nn
=\exp\{\Tr\ln\Dvec\}=\exp[-\Tr
\int_0^\infty\frac{\e^{-t\Dvec}}{t}dt+\mbox{const}\ ]
\label{CF2}
\end{eqnarray}
Clearly the following things are different from Sec.2:\ 
1)\ Dimension of $t$ is length ($\Dvec$ is the {\it linear} differential
operator);\ 
2)\ $\Dvec$ has both positive and negative solutions. Due to 2),
the $t$-integral is divergent and
further regularization is necessary to define $\det \Dvec$.
The purpose of this paper is to define the integrand
inside the $t$-integral above,
using the operator $\Dvec$ in such a way that the $t$-integral is
convergent and the property as the determinant is preserved
(regularization of the fermion determinant). We will do it in three stages.

\flushleft{(1) First Stage}

\q Continuing formally a little longer,
the heat-kernel $G(x,y;t)=<x|e^{-t\Dvec}|y>$ 
satisfies (\ref{HK4}). 
The operator $\exp\{-t\Dvec\}$ satisfies, using $(\gago)^2=1$, 
the relation
\begin{eqnarray}
\exp\{-t\Dvec\}=\half(1+i\ga_5)\exp\{+it\ga_5\Dvec\}+
\half(1-i\ga_5)\exp\{-it\ga_5\Dvec\}              \pr
\label{CF3}
\end{eqnarray}
This is valid for any operators which satisfy
\begin{eqnarray}
\ga_5\Dvec+\Dvec\ga_5=0 \pr
\label{CF4}
\end{eqnarray}
In eq.(\ref{CF3}), $\half (1\pm i\ga_5)$ are Wick-rotated variations of
the chiral projection operators $P_\pm=\half (1\pm \ga_5)$. This
hints at the 5 dimensionality behind the discussion. 

\q Here we introduce two new heat-kernels corresponding to the two terms
in RHS of (\ref{CF3}).
\begin{eqnarray}
G^5_\pm(x,y;t)\equiv <x|\exp\{\pm it\ga_5\Dvec\}|y>\com\q
(\frac{\pl}{\pl t}\mp i\ga_5\Dvec)G^5_\pm(x,y;t)=0 
\pr
\label{CF5}
\end{eqnarray}
Very interestingly, the above heat equations 
satisfy the 1+4 dim
Minkowski Dirac equation after appropriate Wick rotations for $t$.
\begin{eqnarray}
(i\dbslash-M)\GfMp=i\e \Aslash \GfMp\com\q
(X^a)=(-it,x^\m)\com                    \nn
(i\dbslash-M')\GfMpm =i\e \Aslash \GfMpm\com\q
(X^a)=(+it,x^\m)
\com\nn
\GfMp(x,y;t)= <x|\exp\{+it\ga_5(\Dvec+iM)\}|y>\com\nn
\GfMpm(x,y;t)= <x|\exp\{-it\ga_5(\Dvec+iM')\}|y>
\pr
\label{CF6}
\end{eqnarray}
where $\Aslash\equiv\ga_\m A_\m(x)\ ,\ 
\dbslash\equiv\Ga^a\frac{\pl}{\pl X^a}$, and
 we have introduced new regularization mass parameters $M,M'(\ra 0)$
for the next step\cite{foot3}. 
Note that 
the Wick-rotation is taken differently for $\GfMp$ and $\GfMpm$.
The sign convention 
for the 1+4 dim fermion masses formally
follows the textbook by Bjorken and
Drell\cite{BD}. However we do {\it not} fix their signs here. 
In the above Dirac equations, $\Aslash$ 
in the right-hand sides is purely
a 4 dim object\cite{foot4}. 
Others are 1+4 dim ones. Hence we obtain a regularized
partition function, at the present stage, as
\begin{eqnarray}
\ln\,Z=-\int_0^{\infty}\frac{dt}{t}\Tr\left[
\half(1+i\ga_5)\exp\{it\ga_5\Dvec\}+
\half(1-i\ga_5)\exp\{-it\ga_5\Dvec\}\right]\nn
=-\lim_{M\ra 0}\int_0^{\infty}\frac{dt}{t}
\half(1-i\frac{\pl}{\pl(tM)})\Tr \GfMp(x,y;t)  \nn
-\lim_{M'\ra 0}\int_0^{\infty}\frac{dt}{t}
\half(1-i\frac{\pl}{\pl(tM')})\Tr \GfMpm(x,y;t)
  \pr
\label{CF7}
\end{eqnarray}
Here we understand $\GfMp$ and $\GfMpm$ are, first calculated
in the 1+4 dim$(X^a)$, and then go back to $(t,x^\m)$ by the Wick rotations.
Note ``Tr'' means the 4 dimensional trace (not over the 5-th dimension).
The r\^{o}le of $M,M'$ looks just like a technical trick at present. 
From the usage above, the limiting condition
for the parameters ($M,M'\ra 0$) should be taken as
\begin{eqnarray}
t|M|\ll 1\com\q t|M'|\ll 1
  \pr
\label{CF8}
\end{eqnarray}
In the following we take $M=M'$ for simplicity.
\vs {0.5}
\flushleft{(2) Second Stage}

\q Next stage of the regularization program is to give
solutions to $\GfMp$ and $\GfMpm$, 
taking into account the boundary conditions. 
We do the calculation in 1+4 dim. The perturbative
solution of the 1+4 dim Dirac equation is again given by Ref.\cite{BD}. 
\begin{eqnarray}
(i\dbslash-M)G^5_{M}=i\e \Aslash G^5_{M}\com\nn
G^5_M(X,Y)=G_0(X,Y)+\int d^5Z\,S(X,Z)i\e\Aslash(z)G^5_M(Z,Y)\com
\label{CF9}
\end{eqnarray}
where $(X^a)=(X^0,X^\m=x^\m)$ and 
$G_0(X,Y)$ is the free solution and $S(X,Z)$ is the propagator.
\begin{eqnarray}
(i\dbslash-M)G_0(X,Y)=0\com\q
(i\dbslash-M)S(X,Y)=\del^5(X-Y)\pr
\label{CF10}
\end{eqnarray}
Solutions depend on boundary conditions
on $X^0-Y^0$. 
There are four possible ways to go around the positive
and negative energy poles in the $K^0$-integral
in its complex plane. (See Fig.1.)
\begin{figure}
\centerline{\epsfysize=4cm\epsfbox{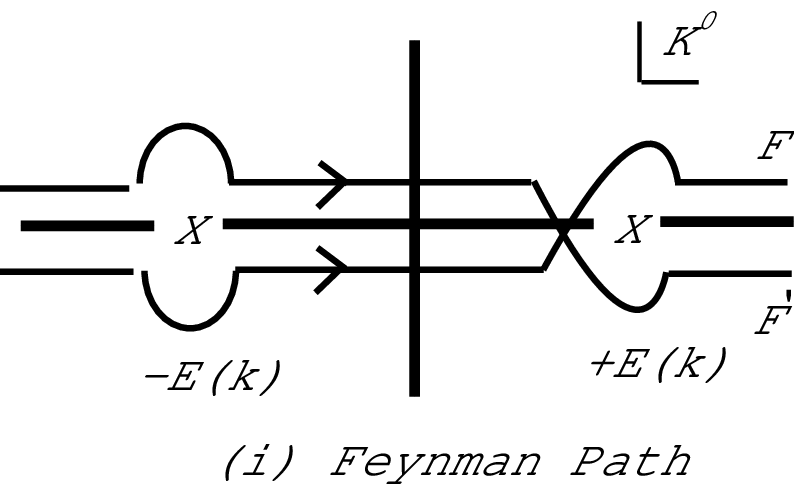}\\
            \epsfysize=4cm\epsfbox{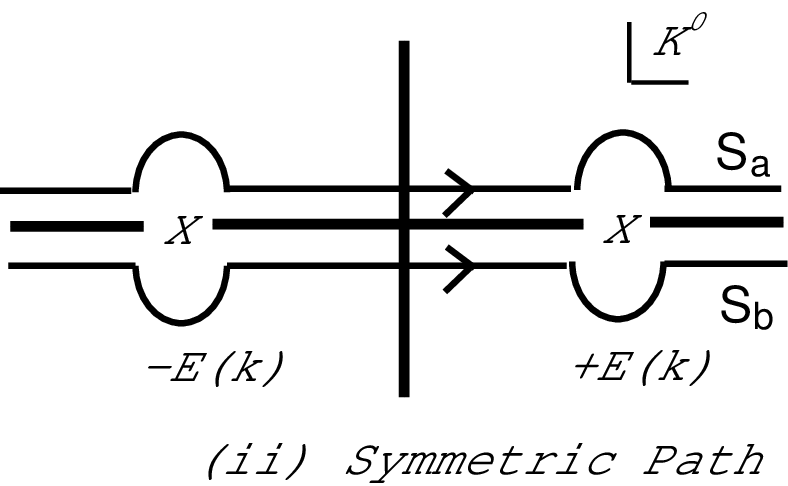}}
   \begin{center}
Fig.1\ 
Four possible pathes for the 1+4 dim Dirac Fermion propagator.
 $E(k)=\sqrt{k^2+M^2}>0.$
   \end{center}
\end{figure}
\ 1)[F-path]\ The familiar Feynman propagator
takes the path:\  ``below'' for the negative pole and ``above''
for the positive one. This treats the positive and negative poles
discriminatively and gives the retarded propagator
 and the advanced propagator. 
\ 2)[F$'$-path]\ Its opposite choice (``above'' for negative and
``below'' for positive) can also be considered.
The other pair of choices treats both poles equally.
\ 3)[Sa-path]\ The path taken as ``above'' for the positive
and negative poles gives a retarded propagator.
\ 4)[Sb-path]\ Its counter choice, ``below'' for both poles, gives 
an advanced propagator. 
Let us discuss them separately.\nl
\flushleft{i)\ Feynman Propagator (F-path,F$'$-path)}\nl
First we consider the Feynman propagator.
\begin{eqnarray}
S_F(X,Y)=\int\frac{d^5K}{(2\pi)^5}
\frac{\e^{-iK (X-Y)}}{\Kbslash-M+i\ep}\com\q
\Kbslash=\Ga^a K_a\com
\label{CF11}
\end{eqnarray}
where $\ep$($\ra +0$) is introduced to define the boundary
condition.
(Note that $\ep$ is only for specifying the path 
of the $K^0$-integral and should not be
regarded as a present regularization parameter.)
After the $K^0$-integral, we see $S_F(X,Y)$ is composed
of the retarded part and the advanced one in $X^0$-axis.
\begin{eqnarray}
S_F(X,Y)=\th (X^0-Y^0)G^p_0(X,Y)+\th(Y^0-X^0)G^n_0(X,Y)\com\nn
G^p_0(X,Y)\equiv -i\int\frac{d^4k}{(2\pi)^4}\Om_+(k)\e^{-i\Ktil(X-Y)}\com\ \ 
\Om_+(k)\equiv\frac{M+\Ktilbslash}{2E(k)}\com\nn
G^n_0(X,Y)\equiv -i\int\frac{d^4k}{(2\pi)^4}\Om_-(k)\e^{+i\Kbar(X-Y)}\com\ \ 
\Om_-(k)\equiv\frac{M-\Kbarbslash}{2E(k)}\com
\label{CF12}
\end{eqnarray}
where $E(k)=\sqrt{k^2+M^2},(\Ktil^a)=(E(k),K^\m=-k^\m),
(\Kbar^a)=(E(k),-K^\m=k^\m)$. 
$k^\m$ is the momentum in the 4 dim Euclidean space.\cite{foot5}
$\Ktil$ and $\Kbar$ are on-shell
momenta($\Ktil^2=\Kbar^2=M^2$), which correspond to the positive
and negative energy states respectively.\cite{foot6}
 Hence $G^p_0(X,Y)$
($G^n_0(X,Y)$) is constructed by the positive (negative) energy
eigenstates (see Ref.\cite{BD}).

\q In analogy to Sec.2, we should take a solution in such a way
that there exists a fixed direction in time. Taking into account 
the $t$-integral convergence, we are uniquely led to
the following solution of $\GfMp$ and $\GfMm$.
\begin{eqnarray}
\mbox{Retarded solution for}\q \GfMp:\q \nn
G_0(X,Y)=G^p_0(X,Y)\com\q S(X,Y)=\th (X^0-Y^0)G^p_0(X,Y)\label{CF13A}\\
\mbox{Advanced solution for}\q \GfMm:\q \nn
G_0(X,Y)=G^n_0(X,Y)\com\q S(X,Y)=\th (Y^0-X^0)G^n_0(X,Y)
\label{CF13B}
\end{eqnarray}
This is the first choice of the 2nd stage regularization. 
The configuration where the positive energy states propagate
only in the forward direction of $X^0$ constitutes the ($+$)-domain,
while the configuration where the negative enegy states propagate 
only in the backward direction constitutes the ($-$)-domain.

\q The similar regularization is obtained by the opposite
choice of path to Feynman propagator, that is, 2)[F$'$-path].

\flushleft{ii)\ Symmetric Propagators (Sa-path,Sb-path)}\nl
We have another choices of the $K^0$-integral path which is
symmetric with respect to positive and negative energy states.
\begin{eqnarray}
S^{\mbox{ret}}_{\mbox{sym}}(X,Y)
=\th(X^0-Y^0)G^{\mbox{p-n}}_0(X,Y)\com                 \nn
G^{\mbox{p-n}}_0(X,Y)\equiv -i\int\frac{d^4k}{(2\pi)^4}
\{ \Om_+(k)\e^{-i\Ktil (X-Y)}-\Om_-(k)\e^{i\Kbar (X-Y)} \}\nn
=G^p_0(X,Y)-G^n_0(X,Y)\com                            \nn
S^{\mbox{adv}}_{\mbox{sym}}(X,Y)
=\th(Y^0-X^0)G^{\mbox{n-p}}_0(X,Y)\com               \nn
G^{\mbox{n-p}}_0(X,Y)\equiv -G^{\mbox{p-n}}_0(X,Y)
\com
\label{CF14}
\end{eqnarray}
where $S^{\mbox{ret}}_{\mbox{sym}}$ and 
$S^{\mbox{adv}}_{\mbox{sym}}$ are obtained by
taking Sa-path and Sb-path respectively. 
Using these solutions we obtain the second choice for
the 2nd stage regularization.
\begin{eqnarray}
\mbox{Symmetric retarded solution for}\q \GfMp:\q \nn
G_0(X,Y)=G^{\mbox{p-n}}_0(X,Y)\com\q 
S(X,Y)=\th (X^0-Y^0)G^{\mbox{p-n}}_0(X,Y)\label{CF15A}\\
\mbox{Symmetric advanced solution for}\q \GfMm:\q \nn
G_0(X,Y)=G^{\mbox{n-p}}_0(X,Y)\com\q 
S(X,Y)=\th (Y^0-X^0)G^{\mbox{n-p}}_0(X,Y)
\label{CF15B}
\end{eqnarray}
In this second choice, both positive and negative states
propagate in the forward direction in + domain, while
in the backward direction in - domain.

\q Differences between the cases i) and ii) 
will be  explained later.
We consider the two cases as different regularizations
generally.

\vs {0.5}
\flushleft{(3) Third Stage}

\q As the final stage of the regularization, we give the condition:
\begin{eqnarray}
|\frac{k^\m}{M}|\ll 1\pr
\label{CF16}
\end{eqnarray}
This condition corresponds to ``fermion zero mode limit''. 
This limit plays the role of selecting a definite chirality.
In fact in this limit ($M>0$ is taken here), 
the Fourier transform of $G^p_0$ and $G^n_0$ is   
\begin{eqnarray}
iG^p_0(k)\sim\Om_+(k)=
\frac{M+\Ktilbslash}{2E(k)}\ra\frac{1+\gago}{2}\ ,\ 
iG^n_0(k)\sim\Om_-(k)=
\frac{M-\Kbarbslash}{2E(k)}\ra\frac{1-\gago}{2}\ ,
\label{CF18}
\end{eqnarray}
which indicates the present regularization clearly selects
both chiralities in the first choice (of the 2nd stage) and
treats them equally in the second choice.

\vspace{1cm}
\q Let us reconsider another condition (\ref{CF8}). The phases
of $G^p_0$ and $G^n_0$ are given by
\begin{eqnarray}
\e^{-i\Ktil X}=\e^{-iE(k)X^0-ikx}\com\q
\e^{i\Kbar X}=\e^{iE(k)X^0-ikx}\pr
\label{CF19}
\end{eqnarray}
As far as the condition (\ref{CF16}) is valid, the condition
(\ref{CF8}) means $|E(k)X^0|=|E(k)t|\ll 1$. The coordinate
$X^0$ becomes irrelevant and the system dominantly works in the 4 dim
$x^\m$-space (Dimensional reduction\cite{foot7}). 
Furthermore, in the case of the first choice (Feynman, F-path), 
both phases above are proportional to $\e^{-E(k)t}\sim\e^{-|M|t}$ 
in the original coordinate $t$
($X^0=-it$ for the former, $X^0=it$ for the latter).
It says
the fermion is localized within a distance $1/|M|$ 
around the origin of the
extra axis of $t$(``wall'' structure). 
As for the second choice (symmetric), 
both $\e^{-|M|t}$ and $\e^{+|M|t}$
modes coexist. Therefore the configuration looks
like that one wall exists around the origin and the other exists
near the boundary($t=\infty$).
For both choices 
we may say the two conditions (\ref{CF8})
and (\ref{CF16}) combined  
give the reduction to the 4 dim theory. 
Considering these two conditions (\ref{CF8}) and (\ref{CF16}), we
obtain
\begin{eqnarray}
|k^\m|\ll |M|\ll \frac{1}{t} \pr
\label{CF17}
\end{eqnarray}
This relation shows the delicacy in taking the limit in the
1+4 dimensional regularization scheme.
It restricts the configuration
to the ultra-violet region ($t\ll |M|^{-1}$)
in the extra space, whereas
to the infra-red (surface) region in the real 4 dim space ($|k^\m|\ll |M|$)
\cite{foot8}.
In the present case we must note that both $k^\m$ and $t$ are
integration variables. In the concrete calculation below, first
the condition $|k^\m|\ll |M|$ is realized by suppressing
the large $|k^\m|$ region, compared to $|M|$, in the $k^\m$-integral.
(See \cite{foot10} for the practical situation.)
After the $k^\m$-integral, $|M|t\ra +0$ limit is taken before
the $t$-integral is performed.
This relation (\ref{CF17}) implies, in lattice, 
$|M|$ should be appropriately chosen depending
on the regularization scale and the considered 
momentum-region of 4 dim fermions.
( In fact, in the lattice simulation, the best-fit value of $M$
($\sim$ a few Gev for the hadron simulation) looks to depend on
the simulation "environment" such as the size of the ordinary-space-axes,
the size of the extra-axis
and the quark mass\cite{BS97PRL,Col99}.
)

\q Using (\ref{CF18}), we can easily read off the boundary conditions
for the free solutions in (\ref{CF13A}),(\ref{CF13B}),(\ref{CF15A}) and
(\ref{CF15B}). They
are equal to corresponding
full solutions by its construction. Then we obtain the following table
of the boundary conditions.

\vs 1
\begin{tabular}{|c|c|c|}
\hline
  & $\GfMp$(Retarded)            & $\GfMm$(Advanced)                            \\
\hline
Limit & $|M|(X^0-Y^0)\ra +0 $ &$|M|(X^0-Y^0)\ra -0$   \\
\hline
Feynman & $-i\int\frac{d^4k}{(2\pi)^4}\Om_+(k)\e^{-ik(x-y)}$    
&   $-i\int\frac{d^4k}{(2\pi)^4}\Om_-(k)\e^{-ik(x-y)} $   \\
\hline
Symmetric & $-i\gago\del^4(x-y)$                &   $+i\gago\del^4(x-y)$   \\
\hline
\multicolumn{3}{c}{\q}                                                 \\
\multicolumn{3}{c}{Table 1\ \ List of boundary conditions.  }\\
\end{tabular}

\vs 1

\q For the Feynman path, the boundary condition above says
\begin{eqnarray}
i\gago (\GfMp(X,Y)-\GfMm(X,Y))\ra \del^4(x-y)\com\q
|M|\cdot |X^0-Y^0|\ra +0
\com
\label{T1eq}
\end{eqnarray}
which will be used later.

\section{Chiral Anomaly in 2 Dim QED}
Let us evaluate the chiral anomaly to check that the present
regularization works correctly. For simplicity, a 2-dim Abelian
gauge model is taken. The previous formulae are all valid by
the replacement:\ $\m$ runs from 1 to 2, 
$\int d^4k/(2\pi)^4\ra\int d^2k/(2\pi)^2$. 
It is known that the chiral anomaly is obtained by the measure
change due to the chiral transformation of $\psi$\cite{KF79,II96}.
\begin{eqnarray}
\del\psi=i\al(x)\gago\psi\com\q
\del\psibar=\psibar (i\al(x)\gago)\com\q |\al(x)|\ll 1\pr
\label{CA2}
\end{eqnarray}
The variation of the Jacobian is formally given by
\begin{eqnarray}
\del\,\ln\,J\equiv\del\,\ln\,\det\,
\frac{\pl(\psibar',\psi')}{\pl(\psibar,\psi)}
=\Tr\{i\al(x)\gago\del^2(x-y)+i\al(x)\gago{\bar \del}^2(x-y)\}
\com
\label{CA3}
\end{eqnarray}
where $\del^2(x-y)$ and ${\bar \del}^2(x-y)$ 
are two delta functions which are not
necessarily regularized in the same way.
Now we regularize $\gago\del^2(x-y)$ by replacing it by
the heat-kernels obtained in Sec.3. We have two different
choices corresponding to i) and ii) of Sec.3.
In this section we take $M>0$.
\flushleft{i)\ Feynman path}\nl
First we consider the case of Feynman propagator. Taking into account
the boundary condition (\ref{T1eq}), we should take
\begin{eqnarray}
\half\,\del\,\ln\,J=\lim_{M\cdot |X^0-Y^0|\ra +0}
\Tr\,i\al(x)i(\GfMp(X,Y)-\GfMm(X,Y))
\pr
\label{CA4}
\end{eqnarray}
Only the 1-st order (with respect to $A_\m$) perturbation contributes.
The first term is evaluated as
\begin{eqnarray}
\GfMp|_A
=\int^{X^0}_0dZ^0\int d^2ZG^p_0(X,Z)i\e \Aslash(z)G^p_0(Z,Y)\nn
=\int^{X^0}_0dZ^0\int d^2z\int\frac{d^2k}{(2\pi)^2}\frac{d^2l}{(2\pi)^2}\nn
\times (-i)\frac{M+\Ktilbslash}{2E(k)}i e \Aslash(z)
(-i)\frac{M+\Ltilbslash}{2E(l)}
\e^{-i\Ktil(X-Z)}\e^{-i\Ltil(Z-Y)}
\pr
\label{CA5}
\end{eqnarray}
Here we have made an important assumption about the extra axis:\ 
the axis is a half line ( not a (straight) line ) like the temperature ($t$) 
axis of Sec.2.\cite{foot9}
Instead of $z$, we take a shifted variable $z'$ ( we do not
change $Z^0$), and expand $A_\m(z)$ around the ``center''
$(x+y)/2$.
\begin{eqnarray}
z=z'+\frac{x+y}{2}\ ,\ 
A_\m(z)=A_\m(\frac{x+y}{2})+\pl_\al A_\m|_{\frac{x+y}{2}}\cdot {z'}^\al
+O({z'}^2)
\pr
\label{CA7}
\end{eqnarray}
The second term, $\pl_\al A_\m$, contributes to the anomaly. 
The final relevant term, after the momentum
integrals taking $X^0$-coordinate($Y^0=0$), turns out to be
\begin{eqnarray}
\Tr\,\al(x)\GfMp|_A\sim
\inttx\,\al(x)
(-\frac{i}{8\pi}) e \ep_\mn\pl_\m A_\n  \nn
\times
MX^0\int^\infty_0ds\frac{s(s^2+2)}{(s^2+1)^{3/2}}\sin (MX^0\sqrt{s^2+1})
\ra
\inttx\,\al(x)\left[-\frac{i}{8\pi} e \ep_\mn\pl_\m A_\n
\right] 
\com
\label{CA8}
\end{eqnarray}
as $MX^0\ra +0$.\cite{foot10}
In this case, we can also obtain the same result in the original
$t$-coordinate($X^0=-it,Y^0=0$).
\begin{eqnarray}
\Tr\,\al(x)\GfMp|_A\sim
\inttx\,\al(x)
(-\frac{i}{8\pi}) e \ep_\mn\pl_\m A_\n\times
Mt\int^\infty_0ds\frac{s(s^2+2)}{(s^2+1)^{3/2}}\e^{-Mt\sqrt{s^2+1}}\nn
\ra
\inttx\,\al(x)\left[-\frac{i}{8\pi} e \ep_\mn\pl_\m A_\n
\right]\q (Mt\ra +0)
\pr
\label{CA9}
\end{eqnarray}
$\GfMm$ is evaluated similarly and gives the same result except
the sign. The final chiral anomaly is
\begin{eqnarray}
\half\frac{\del}{\del\al(x)}\ln\,J=
\half\frac{1}{J}\frac{\del J}{\del \al(x)}=+\frac{i}{4\pi} e \ep_\mn\pl_\m A_\n
\pr
\label{CA10}
\end{eqnarray}
which is the half of the well-known ABJ (U(1) chiral) anomaly.
This result will be commented after the next case.\nl
\flushleft{ii)\ Symmetric path }\nl
The second choice at the stage 2
of Sec.3 gives us another regularization.
From the boundary condition of the symmetric path, in Table 1, 
we obtain,
\begin{eqnarray}
\del\,\ln\,J=
\lim_{M(X^0-Y^0)\ra +0}\Tr\,i\al(x)i\GfMp(X,Y)\nn
+\lim_{M(X^0-Y^0)\ra -0}\Tr\,i\al(x)(-i)\GfMm(X,Y)\nn
=\lim_{M\cdot |X^0-Y^0|\ra +0}\Tr\,i^2\al(x)(\GfMp(X,Y)-\GfMm(X,Y))
\pr
\label{CA11}
\end{eqnarray}
The (+)-domain term is evaluated from
\begin{eqnarray}
\GfMp|_A
=\int^{X^0}_0dZ^0\int d^2Z(G^p_0(X,Z)-G^n_0(X,Z))i e \Aslash(z)
(G^p_0(Z,Y)-G^n_0(Z,Y))             \nn
=\int^{X^0}_0dZ^0\int d^2z\int\frac{d^2k}{(2\pi)^2}\frac{d^2l}{(2\pi)^2}\nn
\times (-i)(\frac{M+\Ktilbslash}{2E(k)}\e^{-iE(k)(X^0-Z^0)}
           -\frac{M-\Kbarbslash}{2E(k)}\e^{iE(k)(X^0-Z^0)})
i e \Aslash(z)                             \nn
\times (-i)(\frac{M+\Ltilbslash}{2E(l)}\e^{-iE(l)(Z^0-Y^0)}
    -\frac{M-\Lbarbslash}{2E(l)}\e^{iE(l)(Z^0-Y^0)})
\e^{-ik(x-z)-il(z-y)}
\pr
\label{CA12}
\end{eqnarray}
As in i), after expanding $A(z)=A(z'+(x+y)/2)$, we are led to
the following one, as the relevant part for the anomaly,
\begin{eqnarray}
\Tr\,\al(x)\GfMp|_A\sim\nn
=\int^{X^0}_0dZ^0\int\frac{d^2k}{(2\pi)^2}
(-i)^2\cdot ie\cdot (-\frac{1}{i})\int d^2x\al(x)\pl_\m A_\n\nn
\times\tr [\frac{i}{4}\gago (\pl_\m\frac{\kslash}{E})
\ga_\n (-2i\sin EX^0)       \nn
-i\pl_\m E\cdot (X^0-Z^0)\cdot \frac{i\gago\kslash}{2E}
\cdot\ga_\n\cdot 2\cos E(X^0-2Z^0)]      \nn
=\inttx\,\al(x)
(-\frac{i}{4\pi}) e \ep_\mn\pl_\m A_\n          \nn
\times
MX^0\int^\infty_0ds
[\frac{s(s^2+2)}{(s^2+1)^{3/2}}+\frac{s^3}{(s^2+1)^{3/2}}]
\sin (MX^0\sqrt{s^2+1})\nn
\ra
\inttx\,\al(x)\left[-\frac{i}{2\pi} e \ep_\mn\pl_\m A_\n
\right]\q (MX^0\ra +0)
\com
\label{CA13}
\end{eqnarray}
where we have done the $k^\m$-integral in $X^0$-coordinate. In comparison
with the previous one, we cannot do the same calculation using
the $t$-coordinate 
because of the apperance of $\e^{+tE(k)}$ factor.
Adding the $\GfMm$ contribution, we obtain
\begin{eqnarray}
\frac{1}{J}\frac{\del J}{\del \al(x)}=+\frac{i}{\pi} e \ep_\mn\pl_\m A_\n
\pr
\label{CA14}
\end{eqnarray}
which is the known ABJ anomaly (
eqs.(4.156),(4.181), and (12.195) of \cite{AAR91} ) 
and is two times of the previous result (\ref{CA10}).  

\vspace{1 cm}
\q The symmetric path gives the correct value, whereas
the Feynman path gives half of it. The discrepancy
comes from the fact that we have taken 
, for the latter case, the "chirally-divided"
propagators, 
$S_F^+(X,Y)\equiv\th (X^0-Y^0)G^p_0(X,Y),\ 
S_F^-(X,Y)\equiv\th(Y^0-X^0)G^n_0(X,Y)$. 
They do not satisfy (\ref{CF10}), but satisfy (2D version
of)
\begin{eqnarray}
(i\dbslash-M)S_F^+(X,Y)=P_+\del^5(X-Y)+O(\frac{1}{M})\com   \nn
(i\dbslash-M)S_F^-(X,Y)=P_-\del^5(X-Y)+O(\frac{1}{M})\pr
\label{CA15}
\end{eqnarray}
Therefore the chiral anomaly computed in i) Feynman path
is not that of the (2D)QED (\ref{CF1}), but is effectively that
of the chiral (2D)QED:
\begin{eqnarray}
\Lcal_{chiral}=\psibar (\ga_\m\pl_\m+i\e P_\pm \Aslash)\psi
\pr
\label{CA16}
\end{eqnarray}
In fact the Feynman result (\ref{CA10}) is consistent
with the minimal case of (14.12) of \cite{AAR91}.
The practical advantage of the Feynman path is its
simplicity in the evaluation.

\section{Conclusion}
\q The main motivation for the present work is to clarify
the real meaning or the 
r{\^o}le of the extra dimension in the lattice formulation.
So far the extra axis has been given a rather obscure meaning
such as "a sophisticated flavour space"\cite{Sha93}.
In the present formalism,
the 5-th (extra) dimension is interpreted as the 
(inverse) temperature. 
For the (Euclidean) boson system which has
an elliptic differential operator, such as the case of
Sec.2, it is easy to treat its canonical ensemble using
the heat equation (\ref{HK4}) and the temperature can be
introduced without any difficulty. 
For the fermion system, however,
the direct use has been considered hard because of the appearance
of the negative eigenvalues.
We have solved the difficulty by {\it generalizing the concept
of the temperature} as the parameter along which the system
evolves in a {\it fixed direction}. 
It is based on the analogy to the thermo-dynamical system. By choosing
the ``directed'' Dirac fermion propagation in the 1+4 Minkowski
space, we define the temperature. 
Some different definitions of temperature appear
depending on the choice of propagators. We understand 
they correspond to different regularizations. 
To "sort"
the fermion propagation with respect to "forward" and
"backward" in the extra (time) axis controls chiral
properties in the fermion determinant evaluation.

\q Two kinds of regularization
naturally appear depending on the choice of solutions:\
Feynman path and symmetric path. 
Although the present treatment for the Feynman path could look 
a little "artificial" in the kinematical viewpoint,
we stress the decomposition into the advanced and the retarded parts
is quite natural from the requirement of {\it fixed direction}
in the system movement. Furthermore we also point out
its practical usefulness of calculational simplicity. 
The relation between Feynman and symmetric tempts us to identify it
with the relation between the consistent and covariant
anomalies\cite{KF79} in the chiral gauge theories. In the latter case,
the chiral anomaly caused by a non-hermitian operator $\Dhat_{chi}$
is the central concern, and, in the evaluation, typically two types
regularization appear. They can be prescribed by
two operators:\ 
$\Dhat_{cons}=\Dhat_{chi}\Dhat_{chi}+\Dhat_{chi}^\dag\Dhat_{chi}^\dag$
and
$\Dhat_{cov}=\Dhat_{chi}\Dhat_{chi}^\dag+\Dhat_{chi}^\dag\Dhat_{chi}$.
$\Dhat_{cons}$ leads to the consistent anomaly and $\Dhat_{cov}$
to the covariant one\cite{KF85,DAMTP9687}. $\Dhat_{cons}$ is composed of
two non-hermitian operators (which are hermite conjugate each other), 
while $\Dhat_{cov}$ are of two hermitian operators.
We understand that the regularization ambiguity produces two different
anomalies from one chiral theory.
In the present case, things go somewhat contrastively. 
The initial concern is the chiral anomaly
caused by the hermitian (QED) operator (\ref{CF1}). We have found typically
two types of path in the evaluation:\ Feynman and symmetric. 
The latter path leads to the chiral anomaly of QED (the starting theory)
which is hermitian, while
the former path leads to that of the chiral QED (\ref{CA16})
which is non-hermitian. It shows
one can analyze not only an initial non-chiral (hermitian) theory
(by choosing the symmetric path) but also
the {\it chiral version} of the initial theory (by choosing Feynman path).
The Feynman (F$'$-path) path gives the chiral (anti-chiral) determinant,
whereas the symmetric one gives the non-chiral (hermitian) determinant.
In spite of the difference in the above two cases, 
we still regard Feynman and symmetric
as two different regularizations of a same result. 
This is because both results
are simply related (just a factor of 2 in the present model)
and essentially the same as far as the chiral anomaly is concerned.

\q It looks that the usual lattice approach corresponds
to the symmetric path, not to the Feynman path. The configuration
image of the former is two "walls", one around the origin($t=0$) and
the other near the boundary ($t=\infty$). This is noted 
below (\ref{CF19}) and
fits the image in lattice. The present formalism
suggests the possible usefulness of the other domain wall configuration
appearing in the Feynman path
:\ one "wall" at the origin. (If we take F$'$-path, the configuration
is one "wall" at the boundary ($t=\infty$).)
In this respect, the present
approach looks to extend other ones known so far.  

\q In the Feynman path, positive energy 5D "electrons" propagate
to the future (in 1+4 space-time) for $G^{5M}_+$, and negative
energy ones propagate to the past for $G^{5M}_-$.
As for the symmetric path, both positive and negative 5D "electrons"
propagate to the future for $G^{5M}_+$, to the past for $G^{5M}_-$.
This situation is schematically drawn in Fig.2 where
the negative energy 5D "electrons" are expressed as 
the positive energy 5D "positrons"
propagating in the opposite direction in time 
(the Dirac's hole theory). In Fig.2 we show the F$'$-path
solution (ib) besides F-path one (ia).
\begin{figure}
\centerline{\epsfysize=6cm\epsfbox{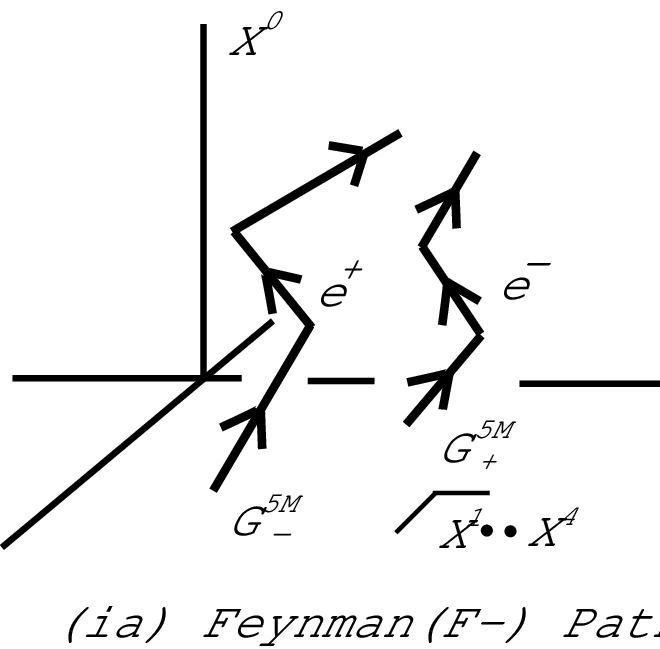}\ \ \epsfysize=6cm\epsfbox{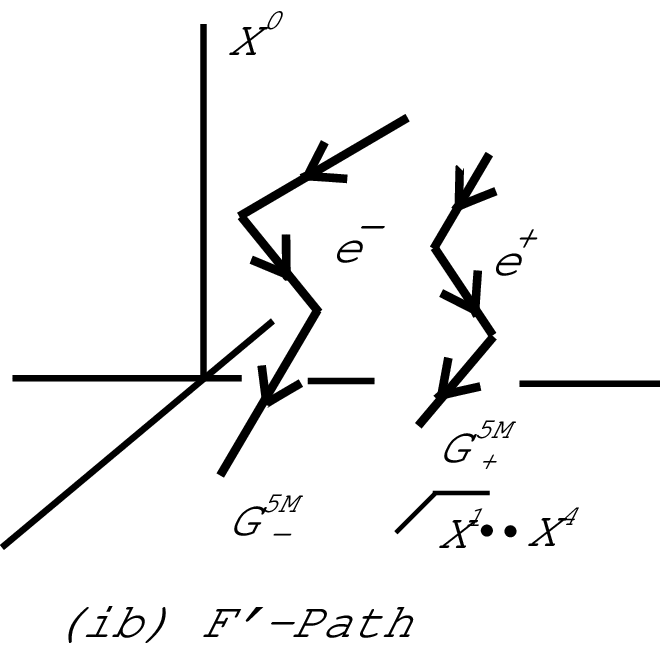}\\
            \epsfysize=6cm\epsfbox{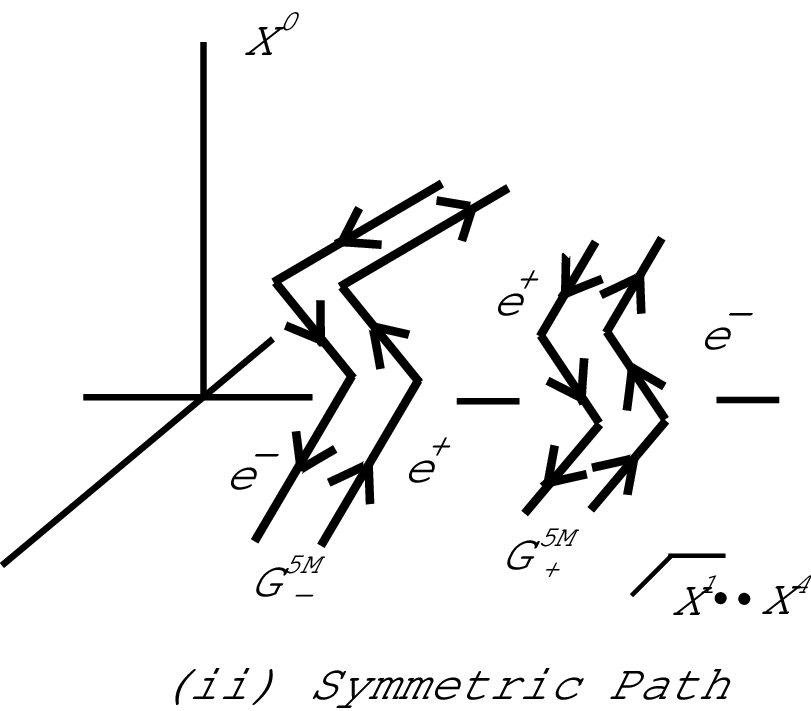}}
   \begin{center}
Fig.2\ 
(+)-domain ($\GfMp$) and ($-$)-domain ($\GfMm$) 
in three different regularizations
:\ (ia) Feynman (F-) path, (ib) F$'$-path and 
(ii) Symmetric path.
$e^-$ represents the positive energy 1+4 dim "electron" and
$e^+$ represents the positive energy 1+4 dim "positron"
which is interpreted as the negative energy "electron"
propagating in the opposite direction in time $X^0$.
   \end{center}
\end{figure}
Note that the present formalism of the chiral determinant
( $\Tr G^{5M}_+$,  $\Tr G^{5M}_-$ ) is based 
not on the vacuum structure but on the solutions of
the (1+4)-dim Dirac equation. This point is different from
the original ones\cite{NN94,NN95,DS95NP,DS95PL,DS97PL}.

\q As for the correspondence to the overlap formalism we point
out some comments.
\begin{enumerate}
\item
$\GfMp$ and $\GfMpm$ correspond to the ``$|+>$ domain''
and ``$|->$ domain'' respectively. 
In the Feynman case, they describe
the right, $(1+\gago)/2$, and the left, $(1-\gago)/2$,
chiral fermion contributions, respectively, 
in the limit of (\ref{CF17}). If we change the sign of the masses,
their chiralities change each other. This corresponds to
the definition of $|\pm>$ in the overlap formalism.
\item
The relations (\ref{CA4}) and (\ref{CA11})
clearly say that the ``$|+>$-domain'' and ``$|->$-domain''
are both necessary to regularize $-i\gago\del^4(x-y)$.
This is the present understanding why the overlap
of $|+>$ and $|->$ 
is necessary to give the correct anomaly.
\end{enumerate}

The results based on our interpretation are consistent
with others known so far. We hope the present approach helps to
further development of chiral fermions
on lattice.

\vs 1
\begin{flushleft}
{\bf Acknowledgment}
\end{flushleft}
The author thanks M. Creutz for continually helping him
at all stages of this work. Without his insight this work
could not be finished.
The author also thanks A. Soni for their latest information about
the lattice simulation using the domain wall. Finally
he thanks the hospitality at the Brookhaven
National Laboratory where this work has been done.
\vs 1



\begin{thebibliography}{99}
\bibitem{foot1} 
{
When we apply this formula in (\ref{HK2}), we must pay
care to zero modes of $\Dhat$. Generally this point is related to
the fixing of local (gauge) symmetry and the regularization
of the infrared behaviour. In the present case, there exists
no problem in this respect. See Chapter IV of \cite{AAR91}.
}
\bibitem{foot2} 
{
We adopt the convention of Ref.\cite{NN95} for the gamma matrices
($\ga_\m$) and the metric in 2 and 4 dim Euclidean space and 
for those ($\Ga_a$) in 1+4 dim Minkowski space. 
}
\bibitem{foot3} 
{
Here we introduce two independent regularization parameters
, $M$ and $M'$, in order to stress that the two domains are
defined not by (the sign of) the 1+4 dim fermion mass but
by the relation (\ref{CF3}). This point distinguishes
the present formalism from the original overlap formalism.
We will soon take $M=M'$ for simplicity.
}
\bibitem{foot4} 
{
A slash '/' is used for 4 dim gamma matrix ($\ga_\m$) contraction, whereas
a backslash '$\backslash$' for 1+4 dim one ($\Ga_a$).  
}
\bibitem{foot7} 
{
In the Kaluza-Klein reduction, the extra dimension is compactified
as $S^1$ ( circle ). In the present case, however, the extra space
can not be compactified since it is known, in the lattice formalism,
that the fermion doubler appears in the case when the periodic
boundary condition is imposed on the extra space\cite{Sha93}. 
}
\bibitem{foot8} 
{
In the standpoint of the 4 dim momentum integral, 
the limit $|M|t\ra +0$ plays the role of the infra-red regularization,
whereas the limit $|M|/|k^\m|\ra +\infty$ that of the ultra-violet
regularization. In the sense that both the ultra-violet and
the infra-red cut-offs exist, 
the present regularization
is similar to the lattice regularization.
}
\bibitem{foot9} 
{
If we take the extra axis as a (straight) line, 
we have to introduce an infrared 
cut-off,$L$, in the $Z^0$-integral in (\ref{CA5}), and the final
anomaly result depends on it.
}
\bibitem{foot10} 
{
Here the momentum ($s=\sqrt{k^2}/M$) integral is
regularized by the analytic continuation. It is 
essentially equivelent
to introduce the cut-off $M$ in the $s$-integral as
$s=\sqrt{k^2}/M\leq 1$. In this sense, the third stage condition
(\ref{CF16}) is used in a loosened way. 
We regard it as a part of regularization.
The same thing can be said for the evaluation of (\ref{CA13}). 
}
\end{thebibliography}
\end{document}